\documentclass[journal]{IEEEtran}
\usepackage{amsmath,amsfonts}
\usepackage{algorithm}
\usepackage{algpseudocode}
\usepackage{array}
\usepackage[caption=false,font=normalsize,labelfont=sf,textfont=sf]{subfig}
\usepackage{textcomp}
\usepackage{stfloats}
\usepackage{url}
\usepackage{verbatim}
\usepackage{graphicx}
\usepackage{cite}
\usepackage{xcolor}
\usepackage{bm}

\usepackage{pgfplots}           %for plotting graphs
\usepackage{graphics}
\usepackage{tikz}
\usepackage{mathdots}
\usepackage{yhmath}
\usepackage{cancel}
\usepackage{color}
\usepackage{siunitx}
\usepackage{array}
\usepackage{multirow}
\usepackage{textcomp}
\usepackage{gensymb}
\usepackage{tabularx}
\usepackage{booktabs}
\usepackage{graphicx}
\usepackage{xcolor}
\pgfplotsset{compat=1.16}

\newtheorem{remark}{Remark}

\newcommand{\bpsi}{\boldsymbol{\psi}}

\begin{document}

\title{Minimum Data Rate Maximization for Uplink Pinching-Antenna Systems}

\author{Sotiris A. Tegos,~\IEEEmembership{Senior Member,~IEEE,} Panagiotis D. Diamantoulakis,~\IEEEmembership{Senior Member,~IEEE,} \\ Zhiguo Ding,~\IEEEmembership{Fellow,~IEEE,} and~George K.~Karagiannidis,~\IEEEmembership{Fellow,~IEEE}% <-this % stops a space

\thanks{S. A. Tegos, P. D. Diamantoulakis, and G. K. Karagiannidis are with the Aristotle University of Thessaloniki, 54124 Thessaloniki, Greece (tegosoti@auth.gr, padiaman@auth.gr, geokarag@auth.gr).}
\thanks{Z. Ding is with the University of Manchester and the Khalifa University, Abu Dhabi, UAE. (email: zhiguo.ding@manchester.ac.uk).}
\thanks{The work was implemented in the framework of HFRI call ``Basic research financing (horizontal support of all sciences)'' under the National Recovery and Resilience Plan ``Greece 2.0'' funded by the European Union - NextGenerationEU (HFRI Project Number: 15642). Z. Ding’s work was supported by the UK EPSRC under grant number EP/W034522/1 and by H2020-MSCA-RISE-2020 under grant number 101006411.}
% \thanks{Manuscript received April 19, 2021; revised August 16, 2021.}
\vspace{-5mm}
}

\maketitle

\begin{abstract}
This paper addresses, for the first time, the uplink performance optimization of multi-user pinching-antenna (PA) systems, recently developed for next-generation wireless networks. By leveraging the unique capabilities of PAs to dynamically configure wireless channels, we focus on maximizing the minimum achievable data rate between devices to achieve a balanced trade-off between throughput and fairness. An effective approach is proposed that separately optimizes the positions of the PAs and the resource allocation. The antenna positioning problem is reformulated into a convex one, while a closed-form solution is provided for the resource allocation. Simulation results demonstrate the superior performance of the investigated system using the proposed algorithm over corresponding counterparts, emphasizing the significant potential of PA systems for robust and efficient uplink communication in next-generation wireless networks.
\end{abstract}


\section{Introduction}
Meeting the increasing connectivity demands of users and internet-of-things (IoT) devices in diverse environments is a critical challenge for the design of next-generation wireless networks. In this new paradigm, wireless communication systems must provide symmetric performance in the downlink and uplink. This is because they are envisioned to be seamlessly integrated with distributed computing frameworks and edge intelligence to enable real-time processing, efficient resource allocation, and context-aware decision making at the network edge, the effectiveness of which is highly dependent on the uplink \cite{proceedings}. However, due to the energy limitations of mobile devices, the achievable data rates in the uplink of wireless communication systems are highly dependent on path loss. Technologies such as centralized multiple-input multiple-output (MIMO) have shown great promise in tackling small-scale fading, thus increasing signal reliability and improving data rates \cite{Zhang2022Beam}, however their performance is still degraded as path loss increases.

Recent innovations in flexible-antenna systems, such as reconfigurable intelligent surfaces (RISs), fluid antennas, and movable antennas, have further pushed the boundaries of wireless communications. These systems dynamically reconfigure wireless channels to improve performance. For example, RISs, acting as relays, use adjustable electromagnetic elements to modify signal characteristics, such as signal amplitudes and phases \cite{Liu2021Reconfigurable, Tegos2022}. While RISs can effectively mitigate line-of-sight (LoS) blockage by reflecting signals around obstacles, they introduce double attenuation caused by the incoming and outgoing channels \cite{Tyrovolas2024}. Similarly, fluid antennas made of reconfigurable materials, such as liquid metals, allow for real-time antenna shape and position adjustments, making them suitable for compact, space-constrained devices \cite{Wong2021Fluid}. Movable antennas, which change their physical position to optimize signal coverage, also provide adaptability to dynamic environments \cite{Zhu2024Modeling}. However, the positional adjustments of fluid and movable antennas occur at the base station (BS) premises and are insufficient to address large-scale path loss, especially at higher carrier frequencies where wavelengths are shorter. It should be noted that under non-LoS conditions, the resulting path loss becomes significantly greater than under LoS conditions, making the establishment of robust LoS links a critical requirement for reliable communications. In addition, existing flexible-antenna systems lack antenna reconfiguration flexibility, i.e., adding or removing an antenna element in these flexible-antenna systems is not straightforward.

To this end, pinching antennas (PAs) can be used, which operate as leaky wave antennas by using small dielectric particles to activate specific points along a waveguide that can be tens of meters long. This technology, first introduced by DOCOMO in 2022 \cite{Suzuki2022Pinching}, enables precise control of radiation locations without the need for additional hardware, a practical and effective approach that ensures antennas can be flexibly deployed in optimal positions \cite{Ding2024Flexible}. Unlike conventional antennas, PAs can cost-effectively and dynamically establish and maintain stable LoS links, even in complex and obstructed environments. This adaptability significantly mitigates the effects of LoS blockage and path loss, ensuring reliable communications with minimal deployment costs. In addition, the delayed taps of a wireless channel, traditionally considered fixed and uncontrollable parameters, can now be actively configured and optimized. 
% In addition, by enabling fine control of communication regions, PAs provide a practical and simple means of mitigating the effects of both LoS blockage and path loss, thus increasing coverage and reliability.

This paper considers, for the first time in the literature, the uplink of multi-user PA systems. To avoid inter-channel interference, orthogonal multiple access (OMA) is used, where a challenging issue is how to allocate the available resources to the devices. Furthermore, to realize the full potential of PA systems, the positions of PAs need to be carefully selected to meet the quality-of-service requirements of multiple devices, which makes the addressed uplink design more challenging than that of conventional systems. Specifically, we formulate an optimization problem that aims at maximizing the minimum data rate between devices to balance the trade-off between system throughput and user fairness. To solve the formulated problem, we separately solve the problem of optimizing the positions of the PAs with fixed power allocation and the resource allocation problem for fixed positions of the PAs. It is shown that the challenging antenna position optimization problem can be transformed into a convex form, which can be solved efficiently. Furthermore, a closed-form solution for the optimal resource allocation problem is obtained by utilizing its convex property. Simulations show the performance of the PA uplink system and highlight the effectiveness of the proposed algorithm over corresponding counterparts.

\begin{figure}
    \centering
\tikzset{every picture/.style={line width=0.7pt}} %set default line width to 0.75pt        

\begin{tikzpicture}[x=0.65pt,y=0.65pt,yscale=-1,xscale=0.95]

%Shape: Parallelogram [id:dp05455380536353094] 
\draw   (141.44,46.11) -- (403.5,46.11) -- (291.19,207.6) -- (29.13,207.6) -- cycle ;
%Straight Lines [id:da909615884514557] 
\draw  [dash pattern={on 4.5pt off 4.5pt}]  (17.75,224.1) -- (158.04,22.06) ;
\draw [shift={(159.75,19.6)}, rotate = 124.78] [fill={rgb, 255:red, 0; green, 0; blue, 0 }  ][line width=0.08]  [draw opacity=0] (8.93,-4.29) -- (0,0) -- (8.93,4.29) -- cycle    ;
%Straight Lines [id:da07314512908880255] 
\draw  [dash pattern={on 4.5pt off 4.5pt}]  (57.38,127.64) -- (351.78,127.64) -- (383,127.64) ;
\draw [shift={(386,127.3)}, rotate = 180.39] [fill={rgb, 255:red, 0; green, 0; blue, 0 }  ][line width=0.08]  [draw opacity=0] (8.93,-4.29) -- (0,0) -- (8.93,4.29) -- cycle    ;
%Shape: Rectangle [id:dp6017765118540368] 
\draw  [fill={rgb, 255:red, 235; green, 232; blue, 233 }  ,fill opacity=1 ] (386.98,65.7) -- (85.66,66.49) -- (85.68,72.59) -- (387,71.8) -- cycle ;
%Straight Lines [id:da2730415037752266] 
\draw  [dash pattern={on 4.5pt off 4.5pt}]  (85.68,72.59) -- (85.38,127.88) ;
%Shape: Triangle [id:dp3606544742427469] 
\draw  [fill={rgb, 255:red, 0; green, 0; blue, 0 }  ,fill opacity=1 ] (232.76,77.06) -- (223.96,62.56) -- (241.28,62.41) -- cycle ;
%Image [id:dp34845591656860675] 
\draw (213.34,168.96) node  {\includegraphics[width=20.16pt,height=18.69pt]{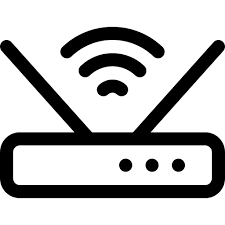}};
%Image [id:dp8327905559643548] 
\draw (27,149.63) node  {\includegraphics[width=31.5pt,height=31.07pt]{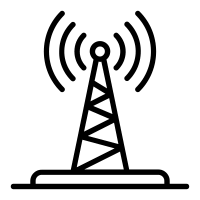}};
%Shape: Triangle [id:dp9506789284994152] 
\draw  [fill={rgb, 255:red, 0; green, 0; blue, 0 }  ,fill opacity=1 ] (324.76,77.06) -- (315.96,62.56) -- (333.28,62.41) -- cycle ;
%Shape: Triangle [id:dp6380202135579276] 
\draw  [fill={rgb, 255:red, 0; green, 0; blue, 0 }  ,fill opacity=1 ] (157.76,77.06) -- (148.96,62.56) -- (166.28,62.41) -- cycle ;
%Image [id:dp4464524109168685] 
\draw (323.34,112.46) node  {\includegraphics[width=20.16pt,height=18.69pt]{Figures/r.png}};
%Image [id:dp6986664828639767] 
\draw (115.34,144.46) node  {\includegraphics[width=20.16pt,height=18.69pt]{Figures/r.png}};

% Text Node
\draw (31,107.99) node [anchor=north west][inner sep=0.75pt]    {$( 0,0,0)$};
% Text Node
\draw (380,110) node [anchor=north west][inner sep=0.75pt]    {$x$};
% Text Node
\draw (138,11.72) node [anchor=north west][inner sep=0.75pt]    {$y$};
% Text Node
\draw (175.9,81.26) node [anchor=north west][inner sep=0.75pt]    {$\boldsymbol{\psi }_{n}^{\text{P}} =\left( x_{n}^{\text{P}} ,0,d\right)$};
% Text Node
\draw (73,86.37) node [anchor=north west][inner sep=0.75pt]    {$d$};
% Text Node
\draw (38.4,46.77) node [anchor=north west][inner sep=0.75pt]   [align=left] {waveguide};
% Text Node
\draw (172.4,45.5) node [anchor=north west][inner sep=0.75pt]   [align=left] {pinching antenna $\displaystyle n$};
% Text Node
\draw (152.7,180.89) node [anchor=north west][inner sep=0.75pt]    {$\boldsymbol{\psi }_{m} =\left( x_{m} ,y_{m} ,0\right)$};

\end{tikzpicture}
    \caption{An illustration for the considered PA uplink system.}
    \label{Fig1}
\end{figure}

\section{System Model}
We consider an uplink communication system consisting of a BS with $N$ PAs in a set $\mathcal{N} = \{1,\dots , N\}$ and $M$ single-antenna devices in a set $\mathcal{M} = \{1,\dots , M\}$. Considering a three-dimensional coordination system, we assume that the devices are randomly deployed in a rectangular area lying in the $x$-$y$ plane with sides $D_x$ and $D_y$, and $\bpsi_m = (x_m, y_m, 0)$ denotes the position of the $m$-th device, as shown in Fig. \ref{Fig1}. Specifically, $x_m$ follows a uniform distribution in $[0,D_x]$ and $y_m$ follows a uniform distribution in $[-D_y/2,D_y/2]$.

In the PA system, it is assumed, without loss of generality, that the waveguide is installed to be parallel to the $x$-axis, with its height denoted by $d$ and its length equal to the side of the rectangular area $D_x$. Thus, the position of the $n$-th PA is $\bpsi_n^{\text{P}} = \left(x_n^{\text{P}},0,d\right)$, where $x_n^{\text{P}} \in [0, D_x]$. The channel between the $m$-th device and the PAs is given by
\begin{equation} \label{c1}
\mathbf{h}_{m,1}=\left[\frac{\sqrt{\eta} e^{-j \frac{2 \pi}{\lambda}\left\|\bpsi_m-\bpsi_1^{\text{P}}\right\|}}{\left\|\bpsi_m-\bpsi_1^{\text {P}}\right\|}, \ldots, \frac{\sqrt{\eta} e^{-j \frac{2 \pi}{\lambda}\left\|\bpsi_m-\bpsi_N^{\text{P}}\right\|}}{\left\|\bpsi_m-\bpsi_N^{\text{P}}\right\|}\right]^{\mathrm{T}} \!\! ,
\end{equation}
where $\eta = c^2/(16 \pi^2 f_c^2)$ is a constant with $c$, $f_c$, and $\lambda$ denoting the speed of light, the carrier frequency, and the wavelength in free space, respectively. 
Since all $N$ PAs are positioned along the same waveguide, the signal transmitted by any antenna is essentially a phase-shifted replica of the signal transmitted by the BS at the feed point of the waveguide \cite{Pozar2021Microwave}, thus the channel in the waveguide is given by
\begin{equation} \label{c2}
    \mathbf{h}_2=\left[e^{-j \frac{2 \pi}{\lambda_g} \left\|\bpsi_0^{\text{P}}-\bpsi_1^{\text{P}}\right\|}, \ldots, e^{-j \frac{2 \pi}{\lambda_g} \left\|\bpsi_0^{\text{P}}-\bpsi_N^{\text{P}}\right\|}\right]^{\mathrm{T}} ,
\end{equation}
where $\psi_0^{\text{P}}$ denotes the position of the feed point of the waveguide, and $\lambda_g = \lambda/n_{\text{e}}$ denotes the guided wavelength with $n_{\text{e}}$ being the effective refractive index of the dielectric waveguide \cite{Pozar2021Microwave}.

The received signal at the BS with the $m$-th device as the transmitter can be written as
\begin{equation}
y_m^{\text{P}}=\sqrt{\frac{P_m}{N}}\mathbf{h}_{m,1}^\mathrm{H}\mathbf{h}_2 s_m +w,
\end{equation}
where $s_m$, $w$, and $P_m$ denote the symbol transmitted by the $m$-th device, the additive white Gaussian noise (AWGN) at the BS with zero mean and variance $\sigma^2$, and the transmit power of the $m$-th device, respectively. It should be noted that $N$ in the denominator appears considering that there is an AWGN including antenna noise and interference in each PA. 

Using \eqref{c1} and \eqref{c2}, the received signal at the BS transmitted by the $m$-th device can be rewritten as
\begin{equation} \label{received}
    y_m^{\text{P}} \!=\! \sqrt{\frac{P_m}{N}} \! \left( \sum_{n=1}^{N} \frac{\sqrt{\eta} e^{ j \frac{2\pi}{\lambda} \left\| \bpsi_m - \bpsi_n^{\text{P}} \right\| - j \frac{2\pi}{\lambda_g}\left\|\bpsi_0^{\text{P}}-\bpsi_n^{\text{P}}\right\| }} {\left\| \bpsi_m - \bpsi_n^{\text{P}} \right\|} \right) s_m + w.
\end{equation}
By changing the positions of the PAs $\bpsi_n^{\text{P}}$, $\forall n \in \mathcal{N}$, the phase shifts in $\mathbf{h}_2$ and the path loss in $\mathbf{h}_{m,1}$ can be configured, which is a unique feature of PA systems. It is assumed that the BS knows the positions of the devices, which can be obtained using conventional localization methods or the recently developed integrated sensing and communication techniques.

Consider that an OMA protocol is used, such as time-division multiple access (TDMA) or frequency-division multiple access (FDMA). Furthermore, to reduce the system complexity, the positions of the PAs cannot be changed when a different device is served.
% We also consider that $P_m = \bar{E}_m/(TBq_m) = E_m/q_m$
We also consider for the transmit signal-to-noise ratio that $P_m/(q_{m1}B \sigma^2) = \bar{E}_m/(q_{m1}B q_{m2}T \sigma^2) = E_m/(q_m \sigma^2)$, where $\bar{E}_m$, $T$, $B$, $E_m$ denote the transmit energy of the $m$-th device, the time frame, the bandwidth, and the normalized transmit energy, respectively. Moreover, $q_{m1} \in [0,1]$ and $q_{m2} \in [0,1]$ denote the portions of the bandwidth and time frame, respectively, allocated to the $m$-th device and $q_m = q_{m1} q_{m2} \in [0,1]$. Thus, the achievable data rate for the $m$-th device is given by
\begin{equation} \label{rate}
    \begin{aligned}
        R_m^{\text{P}} & = q_m \log_2 \Bigg( 1 + \frac{\eta E_m}{N q_m \sigma^2} \\ 
        & \qquad \qquad \left. \times \left\lvert \sum_{n=1}^{N} \frac{ e^{ j \frac{2\pi}{\lambda} \left\| \bpsi_m - \bpsi_n^{\text{P}} \right\| - j \frac{2\pi}{\lambda_g}\left\|\bpsi_0^{\text{P}}-\bpsi_n^{\text{P}}\right\| }} {\left\| \bpsi_m - \bpsi_n^{\text{P}} \right\|} \right\rvert^2 \right) .
    \end{aligned}
\end{equation}

\section{Problem Formulation and Solution}
Although orthogonal resources are used in the considered uplink scenario, the multi-user system parameters are inherently coupled because a single set of antenna positions must be determined to serve all devices. This creates interdependencies among the data rates of the devices, as optimizing antenna placement for one device may potentially degrade the performance of others. Therefore, it is essential to balance these trade-offs to ensure user fairness while maximizing overall system performance.
To this end, in the considered uplink PA system, we focus on maximizing the minimum achievable data rate among the $M$ devices by optimizing the positions of the PAs denoted by $\mathbf{x}^{\text{P}} = [x_1^{\text{P}}, \dots , x_N^{\text{P}}]$ and the resource allocation vector $\mathbf{q} = [q_1, \dots , q_M]$. This optimization problem can be formulated as
\begin{equation} \tag{\textbf{P1}} \label{P1}
    \begin{array}{cl}
    \mathop{\max}\limits_{\mathbf{x}^{\text{P}}, \mathbf{q}}
    & \mathop{\min}\limits_m R_m^{\text{P}} \\
    \textbf{s.t.} 
    & \mathrm{C}_1: | x_n^{\text{P}} - x_{n'}^{\text{P}} | \geq \Delta, \ \forall n, n' \in \mathcal{N}, \ n \neq n', \\
    & \mathrm{C}_2: x_n^{\text{P}} \in [0, D_x], \\
    & \mathrm{C}_3: \sum_{m=1}^M q_m \leq 1, \\
    & \mathrm{C}_4: q_m \geq \delta,
    \end{array}
\end{equation}
where $\mathrm{C}_1$ ensures that the distance between adjacent PAs is greater than or equal to $\Delta$ to avoid coupling effects. Also, $\mathrm{C}_2$ limits the resource blocks allocated to the $M$ devices and in $\mathrm{C}_4$, $\delta \to 0$ to avoid the denominator in \eqref{rate} being zero.

To solve \eqref{P1}, we solve separately the problem of optimizing the positions of the PAs with fixed power allocation and the resource allocation problem for fixed positions of the PAs. The former one is formulated as
\begin{equation} \tag{\textbf{P1.1}} \label{P1.1}
    \begin{array}{cl}
    \mathop{\max}\limits_{\mathbf{x}^{\text{P}}}
    & \mathop{\min}\limits_m R_m^{\text{P}} \\
    \textbf{s.t.} 
    & \mathrm{C}_1, \ \mathrm{C}_2,
    \end{array}
\end{equation}
while the latter is formulated as
\begin{equation} \tag{\textbf{P1.2}} \label{P1.2}
    \begin{array}{cl}
    \mathop{\max}\limits_{\mathbf{q}}
    & \mathop{\min}\limits_m R_m^{\text{P}} \\
    \textbf{s.t.} 
    & \mathrm{C}_3,\ \mathrm{C}_4.
    \end{array}
\end{equation}

\subsection{Optimization of PA Positions}
% For \eqref{P1.2}, we consider that the number of PAs is equal to the number of devices, i.e., $N=M$, we propose that the starting position of each PA is above each device. 
There are two objectives for optimizing the PA positions. One is to ensure that the phases of the free-space channels and the waveguide propagation are aligned, and the other is to ensure that the antenna positions result in favorable path losses. The fact that these two objectives must be achieved for all devices makes this optimization problem particularly challenging. To maximize the achievable data rate of the devices \eqref{P1.1}, the phases of all the PAs should be as close to equal as possible for each device. To do this, we maximize the minimum achievable data rate, considering that the PAs are properly placed so that the phases are perfectly aligned, while introducing a constraint that ensures that the phases are indeed aligned, i.e.,
\begin{equation} \tag{\textbf{P1.1.1}} \label{P1.1.1}
    \begin{array}{cl}
    \mathop{\max}\limits_{\mathbf{x}^{\text{P}},
    \boldsymbol{\theta}}
    & \mathop{\min}\limits_m \bar{q}_m \log_2 \left( 1 + \frac{\eta E_m}{N \bar{q}_m \sigma^2} \left( \sum_{n=1}^{N} \frac{1} {\left\| \bpsi_m - \bpsi_n^{\text{P}} \right\|} \right)^2 \right) \\
    \textbf{s.t.} 
    & \mathrm{C}_1, \ \mathrm{C}_2, \\
    & \mathrm{C}_6: \sum\limits_{n=1}^N w_{mn} \left( f(x_n^{\text{P}},\theta_m) \right)^2 \leq \epsilon, \ \forall m \in \mathcal{M},
    \end{array}
\end{equation}
where $\bar{q}_m$ denotes the starting resource allocation for the $m$-th device in which the resources can be considered equally allocated and
\begin{equation}
f(x_n^{\text{P}},\theta_m) = \frac{2\pi}{\lambda} \sqrt{ \left( x_m - x_n^{\text{P}} \right)^2 + y_m^2 + d^2} - \frac{2\pi}{\lambda_g} x_n^{\text{P}} - \theta_m.
\end{equation}
In $\mathrm{C}_6$, the sum of the weighted squared errors with weights $w_{mn}$ between each exponent in \eqref{received} and a constant phase $\theta_m \in \boldsymbol{\theta}$ for the $m$-th device is bounded by a constant $\epsilon$. To account for the modulo-$2\pi$ nature of the phases, the optimization variable $\theta_m$ is unbounded so that it can be equal to the phase term in \eqref{received}. The weights are a convex function of the distance between the the $m$-th device and the starting positions of the $n$-th PA $\bar{\bpsi}_n^{\text{P}}$, where each one is placed above each device, considering that their number is equal $M=N$. Specifically, to achieve the desired user fairness while reducing path loss, we define them as
\begin{equation} \label{weights}
    w_{mn} = \frac{1} {\left\| \bpsi_m - \bar{\bpsi}_n^{\text{P}} \right\|^3}.
\end{equation}
If $M\neq N$, the optimization framework remains the same, except that (7) must be modified appropriately.

To solve \eqref{P1.1.1}, we use an auxiliary optimization variable $\mathcal{R}_1$ that bounds the objective function and consider, without loss of generality, that the position of the feed point of the waveguide is $\psi_0^{\text{P}} = (0,0,d)$, which can be achieved with an appropriate phase delay. We also consider that the PAs are placed in a successive order without loss of generality, which makes $\mathrm{C}_1$ linear. Moreover, for $\mathrm{C}_6$, which is also a non-convex constraint, we introduce a set of auxiliary optimization variables $z_{mn}\in\mathbf{z}$. Therefore, \eqref{P1.1.1} can be rewritten as
\begin{equation} \tag{\textbf{P1.1.2}} \label{P1.1.2}
    \begin{array}{cl}
    \mathop{\max}\limits_{\mathbf{x}^{\text{P}}, \boldsymbol{\theta}, \mathcal{R}_1}
    & \mathcal{R}_1 \\
    \textbf{s.t.} 
    & \mathrm{C}_1': x_n^{\text{P}} - x_{n-1}^{\text{P}} \geq \Delta, \ \forall n, n-1 \in \mathcal{N}, \\
    & \mathrm{C}_2, \\
    & \mathrm{C}_6': \sum\limits_{n=1}^N w_{mn} z_{mn} \leq \epsilon, \ \forall m \in \mathcal{M}, \\
    & \mathrm{C}_7: \sum\limits_{n=1}^{N} \frac{1} {\left\| \bpsi_m - \bpsi_n^{\text{P}} \right\|} \geq D_m(\mathcal{R}_1), \ \forall m \in \mathcal{M}, \\
    & \mathrm{C}_8:  \left( f(x_n^{\text{P}},\theta_m) \right)^2 \leq z_{mn}, \ \forall m \in \mathcal{M}, \ \forall n \in \mathcal{N},
    \end{array}
\end{equation}
where 
\begin{equation}
    D_m(\mathcal{R}_1) = \left(\left(2^{\frac{\mathcal{R}_1}{\bar{q}_m}}-1\right) \frac{N \bar{q}_m \sigma^2}{\eta E_m}\right)^{\frac{1}{2}}.
\end{equation}

In \eqref{P1.1.2}, $\mathrm{C}_6'$ is a linear constraint, while $\mathrm{C}_7$ and $\mathrm{C}_8$ are non-convex constraints. For $\mathrm{C}_7$, we introduce an auxiliary optimization variable $t$ as the objective function without loss of generality since $D_m(\mathcal{R}_1)$ is an increasing function with respect to $\mathcal{R}_1$. This maximizes the left-hand side of $\mathrm{C}_7$, which effectively maximizes the corresponding achievable data rate of the $m$-th device for fixed resource allocation. Furthermore, we use a set of auxiliary optimization variables $v_{mn}\in\mathbf{v}$ for the left-hand side of $\mathrm{C}_7$.
Moreover, we use second-order cone programming for $\mathrm{C}_8$ and \eqref{P1.1.2} is reformulated as
\begin{equation} \tag{\textbf{P1.1.3}} \label{P1.1.3}
    \begin{array}{cl}
    \mathop{\max}\limits_{\mathbf{x}^{\text{P}}, \boldsymbol{\theta}, \mathbf{z}, t, \mathbf{v}}
    & t \\
    \textbf{s.t.} 
    & \mathrm{C}_1',\ \mathrm{C}_2, \ \mathrm{C}_6', \\
    & \mathrm{C}_7': \sum\limits_{n=1}^N v_{mn} \geq t, \ \forall m \in \mathcal{M}, \\
    & \mathrm{C}_8': \|f(x_n^{\text{P}}, \theta_m), \sqrt{z_{mn}}\| \leq \sqrt{2z_{mn}}, \\
    & \qquad \qquad \forall m \in \mathcal{M}, \forall n \in \mathcal{N}, \\
    & \mathrm{C}_9: \left\| \bpsi_m - \bpsi_n^{\text{P}} \right\| \leq \frac{1}{v_{mn}}, \ \forall m \in \mathcal{M}, \ \forall n \in \mathcal{N}, \\ 
    & \mathrm{C}_{10}: v_{mn} \geq 0, \ \forall m \in \mathcal{M}, \ \forall n \in \mathcal{N},    
    \end{array}
\end{equation}
where $\mathrm{C}_7'$ is a linear constraint and $\mathrm{C}_8'$ is convex since the norm in its left-hand side is convex and the square root in its right-hand side is concave. Furthermore, in $\mathrm{C}_9$, the left-hand side is convex as a sum of norms. For its right-hand side, which is non-convex, we use its first-order Taylor approximation around initial points $v_{mn}^0$ and successive convex approximation (SCA) to update these initial points.
Therefore, \eqref{P1.1.3} is finally rewritten in a convex form as 
\begin{equation} \tag{\textbf{P1.1.4}} \label{P1.1.4}
    \begin{array}{cl}
    \mathop{\min}\limits_{\mathbf{x}^{\text{P}}, \boldsymbol{\theta}, \mathbf{z}, t, \mathbf{v}}
    & t \\
    \textbf{s.t.} 
    & \mathrm{C}_1',\ \mathrm{C}_2, \ \mathrm{C}_6', \ \mathrm{C}_7', \ \mathrm{C}_8', \mathrm{C}_{10}, \\
    & \mathrm{C}_9': \left\| \bpsi_m - \bpsi_n^{\text{P}} \right\| \leq \frac{2}{v_{mn}^0} - \frac{v_{mn}}{(v_{mn}^0)^2} , \ \forall m \in \mathcal{M}.
    \end{array}
\end{equation}

\subsection{Resource Allocation}
To solve \eqref{P1.2}, we introduce an auxiliary optimization variable $\mathcal{R}_2$, so that it is reformulated as
\begin{equation} \tag{\textbf{P1.2.1}} \label{P1.2.1}
    \begin{array}{cl}
    \mathop{\max}\limits_{\mathbf{q},\mathcal{R}_2}
    & \mathcal{R}_2 \\
    \textbf{s.t.} 
    & \mathrm{C}_3,\ \mathrm{C}_4, \\
    & \mathrm{C}_5: R_m^{\text{P}} \geq \mathcal{R}_2, \ \forall m \in \mathcal{M},
    \end{array}
\end{equation}
which is convex, since the objective function, $\mathrm{C}_3$ and $\mathrm{C}_4$ are linear and the second derivative of \eqref{rate} with respect to $q_m$ is negative for $q_m \geq 0$.
The Lagrangian function of \eqref{P1.2.1} is given by
\begin{equation} \label{lagrangian}
    \mathcal{L} = - \mathcal{R} - \lambda_1 \left( \sum_{m=1}^M q_m - 1 \right) - \lambda_2 \left( t - R_m^{\text{P}} \right),
\end{equation}
where $\lambda_1, \lambda_2 \geq 0$ are the Lagrange multipliers. Setting the first derivative of \eqref{lagrangian} with respect to $q_m$ equal to $0$, the optimal allocation for the $m$-th device is given as
\begin{equation} \label{qm_opt}
    q_m^* \!=\! \frac{\eta E_m \bar{W}}{N \sigma^2 \! \left( 1 \!+\! \bar{W} \right)} \! \left\lvert \sum_{n=1}^{N} \! \frac{ e^{ j \frac{2\pi}{\lambda} \left\| \bpsi_m - \bpsi_n^{\text{P}} \right\| - j \frac{2\pi}{\lambda_g}\left\|\bpsi_0^{\text{P}}-\bpsi_n^{\text{P}}\right\| }} {\left\| \bpsi_m - \bpsi_n^{\text{P}} \right\|} \right\rvert^2 \!\!\!,
\end{equation}
where $\bar{W} = W \left( -e^{-1-\frac{\lambda_1 \ln2}{\lambda_2}} \right)$ with $W(\cdot)$ being the Lambert W function or product logarithm.

The procedure for solving \eqref{P1} is described in Algorithm~\ref{Alg}. Specifically, the algorithm decouples the optimization problem into two subproblems and solves them sequentially. First, for a range of squared error bounds $\epsilon \in [\epsilon_{\min}, \epsilon_{\max}]$, the PA positions $\mathbf{x}^\text{P}$ are optimized using SCA. At each iteration of SCA, the non-convex constraint $\mathrm{C}_9$ in the antenna positioning subproblem is linearized by a first-order Taylor approximation. The optimization variables are updated iteratively for $K_{\text{SCA}}$ SCA iterations. After determining the PA positions, the minimum achievable rate for the devices is evaluated, and the best positions are stored as the rate improves. Finally, the resource allocation vector $\mathbf{q}$ is optimized in closed form, ensuring fairness among devices. The output of the algorithm includes the optimized PA positions $\mathbf{x}^\text{P}$ and the resource allocation vector $\mathbf{q}$.
The complexity of Algorithm \ref{Alg} is $\mathcal{O}\left(N^3\right)$, which is due to solving \eqref{P1.1.4} using an interior-point method and considering that the complexity of solving \eqref{P1.2.1} is negligible using the closed-form solution in \eqref{qm_opt}. Moreover, Algorithm \ref{Alg} always converges, since the nature of SCA ensures it.

\begin{remark}
    The positions of PAs are assumed to be fixed when serving the $M$ devices. It should be noted that in the case of TDMA, the positions of the PAs in each slot can be configured instantaneously and placed over each device. However, in practice, this instantaneous configuration of the antenna positions can lead to high system complexity and hardware cost, making the formulated optimization problem a practical solution. It is also worth pointing out that in the case of FDMA instantaneous configuration is not possible, since the devices are served simultaneously.
\end{remark}

\begin{algorithm}
\caption{Solution of \eqref{P1}.}
\label{Alg}
\begin{algorithmic}[1]
\Require Number of devices $M$, number of PAs $N$, area $D_x\times D_y$, initial positions $\bpsi_m$ and $\bpsi_n^{\text{P}}$, and system parameters $\eta, \lambda, \lambda_g, \Delta$.

% \State \textbf{Step 1: Solve \eqref{P1.1.4}}

\State Initialize $\bar{\mathbf{q}} = \frac{1}{M} \mathbf{1}_M$, $\mathbf{x}^{\text{P}}$, and $R_{\text{bound}} = -\infty$.

\For{$\epsilon \in  [\epsilon_{\min}, \epsilon_{\max}]$}

    \State Initialize auxiliary variables $v_{mn}^0$ for all $m \in \mathcal{M}$, $n \in \mathcal{N}$.
    % \State \textbf{Step 1.1: Use SCA}
    \For{$k = 1$ to $K_{\text{SCA}}$} 
        \State Use first-order Taylor approximation around $v_{mn}^0$.
        \State Solve \eqref{P1.1.4}.
        \State Update $v_{mn}^0 \gets v_{mn}$.
        % \If{improvement in $t$ is below tolerance} \Comment{Convergence check}
        %     \State 	extbf{Break}
        % \EndIf
    \EndFor
    \State Update PA positions $\mathbf{x}^{\text{P}}$.

    % \State \textbf{Step 1.2: Evaluate minimum data rate for current~$\epsilon$}
    \State Compute $\mathop{\min}\limits_m R_m^{\text{P}}$ using \eqref{rate} for current $\epsilon$ and  $\mathbf{x}^{\text{P}}$.
    \If{$\mathop{\min}\limits_m R_m^{\text{P}} > R_{\text{bound}}$}
    \State Update $R_{\text{bound}} \gets \mathop{\min}\limits_m R_m^{\text{P}}$.
    \State Store current $\mathbf{x}^{\text{P}}$.
    \EndIf
\EndFor
    
% \State \textbf{Step 3: Solve \eqref{P1.2.1}}
\State Solve \eqref{P1.2.1}.

\State \textbf{Output:} Optimized PA positions $\mathbf{x}^{\text{P}}$ and resource allocation vector $\mathbf{q}$.
\end{algorithmic}
\end{algorithm}

\section{Numerical Results}
This section evaluates the performance of the PA systems and the proposed algorithm through simulations. For consistency, the system parameters are chosen as in \cite{Ding2024Flexible}, where the noise power is $-90$ dBm, the carrier frequency $f_c=28$ GHz, the spacing $\Delta=\lambda/2$, the effective refractive index $n_{\text{e}}=1.4$ and $\epsilon_{\min}=0.1$, $\epsilon_{\max}=0.5$. We also consider an area $D_x \times D_y = 30\times 10$ m$^2$. The proposed optimization framework (a) is compared with a corresponding system (b), where we use the optimized positions of the PAs without resource allocation. Specifically, for this scheme, we solve \eqref{P1.1.4} considering that resources are equally allocated and we do not solve \eqref{P1.2.1}. We also compare with a conventional system (c), where the BS is located in the edge of the rectangular area at $(0,0,0)$ and its antennas are placed at $\bpsi_n^{\text{C}} = \left(x_n^{\text{C}},0,d\right)$. The spacing between adjacent antennas is set to $\Delta$ to avoid antenna coupling. Based on the spherical wave channel model \cite{Zhang2022Beam}, the channel vector between the fixed antennas and the $m$-th device is given by
\begin{equation} 
\mathbf{h}_{m}^{\text{C}} \! = \! \left[\! \frac{\sqrt{\eta} e^{-j \frac{2 \pi}{\lambda}\left\|\bpsi_m-\bpsi_1^{\text{C}}\right\|}}{\left\|\bpsi_m-\bpsi_1^{\text{C}}\right\|}, \ldots, \frac{\sqrt{\eta} e^{-j \frac{2 \pi}{\lambda}\left\|\bpsi_m-\bpsi_N^{\text{C}}\right\|}}{\left\|\bpsi_m-\bpsi_N^{\text{C}}\right\|}  \right]^{\mathrm{T}}\!\!\!,
\end{equation}
The achievable data rate in the conventional system is expressed as
\begin{equation}
    R_m^{\text{C}} = q_m \log_2 \left( 1 + \frac{E_m \|\mathbf{h}_m^{\text{C}}\|^2}{q_m \sigma^2} \right).
\end{equation}

\begin{figure}
\centering
\begin{tikzpicture}
   \begin{axis}[
   width = 0.99\linewidth,
   xlabel = {Transmit power (dBm)},
   ylabel = {Maximum min data rate (bits/s/Hz)},
   ymin = 1,
   ymax = 7,
   xmin = 0,
   xmax = 20,
   ytick = {0,1,...,7},
   grid = major,
   legend entries = {{$M=N=2$, (a)},{$M=N=2$, (b)},{$M=N=2$, (c)},{$M=N=3$, (a)},{$M=N=3$, (b)}, {$M=N=3$, (c)},{$M=N=4$, (a)},{$M=N=4$, (b)}, {$M=N=4$, (c)}},
   legend cell align = {left},
   legend style = {font = \tiny},
   % legend pos = north west
   legend style={at={(0,1)},anchor=north west}
   ]
   \addplot[
    blue,
    mark = square,
    mark repeat = 2,
    mark size = 3,
    mark phase = 0,
    line width = 1pt
    ]
    table {Data/Rmin_vs_Pt_h3/Rmin_vs_Pt_D30_MN2.dat};
    \addplot[
    red,
    mark = square,
    mark repeat = 2,
    mark size = 3,
    mark phase = 0,
    mark options = solid, 
    dashed,
    line width = 1pt
    ]
    table {Data/Rmin_vs_Pt_h3/Rmin_vs_Pt_D30_MN2_b.dat};
    \addplot[
    green,
    mark = square,
    mark repeat = 2,
    mark size = 3,
    mark phase = 0,
    mark options = solid, 
    dash dot,
    line width = 1pt
    ]
    table {Data/Rmin_vs_Pt_h3/Rmin_vs_Pt_D30_MN2_c.dat};
    \addplot[
    blue,
    mark = o,
    mark repeat = 2,
    mark size = 3,
    mark phase = 0,
    mark options = solid, 
    solid,
    line width = 1pt
    ]
    table {Data/Rmin_vs_Pt_h3/Rmin_vs_Pt_D30_MN3.dat};
    \addplot[
    red,
    mark = o,
    mark repeat = 2,
    mark size = 3,
    mark phase = 0,
    mark options = solid, 
    dashed,
    line width = 1pt
    ]
    table {Data/Rmin_vs_Pt_h3/Rmin_vs_Pt_D30_MN3_b.dat}; 
    \addplot[
    green,
    mark = o,
    mark repeat = 2,
    mark size = 3,
    mark phase = 0,
    mark options = solid, 
    dash dot,
    line width = 1pt
    ]
    table {Data/Rmin_vs_Pt_h3/Rmin_vs_Pt_D30_MN3_c.dat};
    \addplot[
    blue,
    mark = triangle,
    mark repeat = 2,
    mark size = 3,
    mark phase = 0,
    mark options = solid, 
    solid,
    line width = 1pt
    ]
    table {Data/Rmin_vs_Pt_h3/Rmin_vs_Pt_D30_MN4.dat};
    \addplot[
    red,
    mark = triangle,
    mark repeat = 2,
    mark size = 3,
    mark phase = 0,
    mark options = solid, 
    dashed,
    line width = 1pt
    ]
    table {Data/Rmin_vs_Pt_h3/Rmin_vs_Pt_D30_MN4_b.dat}; 
    \addplot[
    green,
    mark = triangle,
    mark repeat = 2,
    mark size = 3,
    mark phase = 0,
    mark options = solid, 
    dash dot,
    line width = 1pt
    ]
    table {Data/Rmin_vs_Pt_h3/Rmin_vs_Pt_D30_MN4_c.dat};    
    \end{axis}
\end{tikzpicture}
\caption{Maximum minimum achievable data rate versus transmit power for different numbers of devices and PAs and height $d=3$ m.}
\label{Fig2}
\end{figure}
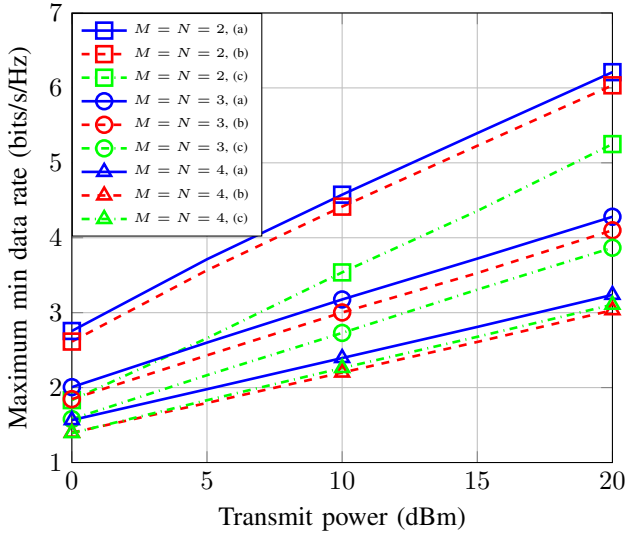

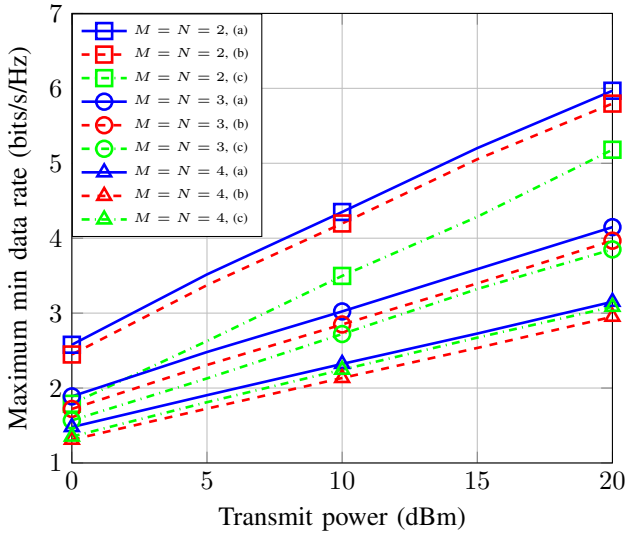
\begin{figure}
\centering
\begin{tikzpicture}
   \begin{axis}[
   width = 0.99\linewidth,
   xlabel = {Transmit power (dBm)},
   ylabel = {Maximum min data rate (bits/s/Hz)},
   ymin = 1,
   ymax = 7,
   xmin = 0,
   xmax = 20,
   ytick = {0,1,...,7},
   grid = major,
   legend entries = {{$M=N=2$, (a)},{$M=N=2$, (b)},{$M=N=2$, (c)},{$M=N=3$, (a)},{$M=N=3$, (b)}, {$M=N=3$, (c)},{$M=N=4$, (a)},{$M=N=4$, (b)}, {$M=N=4$, (c)}},
   legend cell align = {left},
   legend style = {font = \tiny},
   % legend pos = north west
   legend style={at={(0,1)},anchor=north west}
   ]
   \addplot[
    blue,
    mark = square,
    mark repeat = 2,
    mark size = 3,
    mark phase = 0,
    line width = 1pt
    ]
    table {Data/Rmin_vs_Pt_h4/Rmin_vs_Pt_D30_MN2.dat};
    \addplot[
    red,
    mark = square,
    mark repeat = 2,
    mark size = 3,
    mark phase = 0,
    mark options = solid, 
    dashed,
    line width = 1pt
    ]
    table {Data/Rmin_vs_Pt_h4/Rmin_vs_Pt_D30_MN2_b.dat};
    \addplot[
    green,
    mark = square,
    mark repeat = 2,
    mark size = 3,
    mark phase = 0,
    mark options = solid, 
    dash dot,
    line width = 1pt
    ]
    table {Data/Rmin_vs_Pt_h4/Rmin_vs_Pt_D30_MN2_c.dat};
    \addplot[
    blue,
    mark = o,
    mark repeat = 2,
    mark size = 3,
    mark phase = 0,
    mark options = solid, 
    solid,
    line width = 1pt
    ]
    table {Data/Rmin_vs_Pt_h4/Rmin_vs_Pt_D30_MN3.dat};
    \addplot[
    red,
    mark = o,
    mark repeat = 2,
    mark size = 3,
    mark phase = 0,
    mark options = solid, 
    dashed,
    line width = 1pt
    ]
    table {Data/Rmin_vs_Pt_h4/Rmin_vs_Pt_D30_MN3_b.dat}; 
    \addplot[
    green,
    mark = o,
    mark repeat = 2,
    mark size = 3,
    mark phase = 0,
    mark options = solid, 
    dash dot,
    line width = 1pt
    ]
    table {Data/Rmin_vs_Pt_h4/Rmin_vs_Pt_D30_MN3_c.dat};    
    \addplot[
    blue,
    mark = triangle,
    mark repeat = 2,
    mark size = 3,
    mark phase = 0,
    mark options = solid, 
    solid,
    line width = 1pt
    ]
    table {Data/Rmin_vs_Pt_h4/Rmin_vs_Pt_D30_MN4.dat};
    \addplot[
    red,
    mark = triangle,
    mark repeat = 2,
    mark size = 3,
    mark phase = 0,
    mark options = solid, 
    dashed,
    line width = 1pt
    ]
    table {Data/Rmin_vs_Pt_h4/Rmin_vs_Pt_D30_MN4_b.dat}; 
    \addplot[
    green,
    mark = triangle,
    mark repeat = 2,
    mark size = 3,
    mark phase = 0,
    mark options = solid, 
    dash dot,
    line width = 1pt
    ]
    table {Data/Rmin_vs_Pt_h4/Rmin_vs_Pt_D30_MN4_c.dat};    
    \end{axis}
\end{tikzpicture}
\caption{Maximum minimum achievable data rate versus transmit power for different numbers of devices and PAs and height $d=4$ m.}
\label{Fig3}
\end{figure}

Figs. \ref{Fig2} and \ref{Fig3} illustrate the effect of transmit power on the maximum minimum achievable data rate for different configurations of devices and PAs, considering two heights $d = 3$ m and $d = 4$ m. For all configurations, the achievable data rate increases with transmit power. Systems with fewer devices and antennas have steeper slopes, indicating that they benefit more from increases in transmit power than denser systems, which have relatively flatter slopes. Both PA systems consistently outperform the conventional system, confirming the effectiveness of flexible antenna placement in achieving superior performance. In particular, the proposed method achieves significant performance gains over the conventional system, with more than 5 dB gain for $M = N = 2$ and approximately 3 dB gain for $M = N = 3$. These results highlight the importance of optimizing the positions of the PAs, which plays a more critical role than resource allocation in determining system performance. While resource allocation ensures fairness and efficient distribution of resources, dynamic placement of PAs has a greater impact on reducing path loss and aligning with favorable communication channels, thus maximizing achievable data rates. Regarding $M = N = 4$, it can be seen that the gain becomes more limited as $M$ increases, due to the increasing difficulty of aligning the phases of all devices simultaneously. Furthermore, the performance improvement remains evident at both heights considered. However, a slight reduction in achievable rates is observed at the larger height $d = 4$ m, which is attributed to the increased path loss due to the increased distance between the devices and the antennas. 
%These results highlight the robustness and efficiency of the proposed optimization framework.

\section{Conclusions}
This work investigated the uplink performance of PA systems and presented a novel optimization framework aimed at maximizing the minimum achievable data rate across devices to ensure fairness. By utilizing the unique ability of PAs to dynamically reconfigure wireless channels, the study demonstrated significant improvements in addressing challenges such as LoS blockage and large-scale path loss. The proposed approach effectively decouples antenna positioning and resource allocation problems, achieving the desired user fairness. Simulations validated the effectiveness of the methodology and highlighted its superiority over corresponding counterparts. This innovative technology paves the way for practical and scalable solutions to meet the requirements of next-generation communication systems.

\bibliographystyle{IEEEtran}
\bibliography{bib.bib}

\end{document}